\documentclass{article}
\usepackage{epsfig}

\begin{document}

\title{\Large \bf The effect of NLO conformal spins in azimuthal angle decorrelation of jet pairs}
\author{{Agust{\' \i}n Sabio Vera}
\\[2.5ex] {\it Physics Department, Theory Division, CERN,}\\
{\it CH--1211 Geneva 23, Switzerland}}

\maketitle

\vspace{-8cm}
\begin{flushright}
{\small CERN--PH--TH/2006--030}
\end{flushright}

\vspace{7cm}
\begin{abstract}
Azimuthal angle decorrelation in inclusive dijet cross sections is studied 
analytically to take into account the next--to--leading corrections to the 
BFKL kernel while keeping the jet vertices at leading order. The spectral 
representation on the basis of leading order eigenfunctions is generalized 
to include the dependence on conformal spins. With this procedure running 
coupling effects and angular dependences are both included. It is shown how 
the angular decorrelation for jets with a wide relative separation in rapidity 
largely decreases at this higher order in the resummation.
\end{abstract}

\section{Introduction}
Among the many relevant questions still open in Quantum Chromodynamics a very 
interesting one is how to describe scattering amplitudes in the so--called 
Regge limit. In Regge asymptotic the center--of--mass energy, $s$, is much 
larger than all other Mandelstam invariants and mass scales present in the 
process under investigation. It is possible to perturbatively keep track of 
the different contributions to the amplitude if some hard scale is present so 
that the strong coupling remains small. It is then needed to resum 
logarithmically enhanced contributions of the form 
$\left(\alpha_s \ln{s}\right)^n$ to all orders. This is achieved using the 
leading--logarithmic (LL) Balistky--Fadin--Kuraev--Lipatov (BFKL) evolution 
equation~\cite{FKL}.

The BFKL approach predicts a power--like rise in $s$ of total cross sections. 
A lot of attention has been given to the search for BFKL effects in deep 
inelastic scattering (DIS) due to the rapid growth of structure functions at 
small values of Bjorken $x$. However, the golden process where the resummation 
of $\ln{s}$ is most important is the total cross section of two  
photons with large and similar virtualities. In this configuration the small 
$x$ resummation, which includes ordering in rapidity and not in transverse 
scales, should be the dominant contribution to the scattering. This is not 
necessarily the case in DIS where ordering in $k_t$ is important given that 
parton evolution takes place between an object with large transverse size, the 
proton, and a small highly virtual photon.

Observables where BFKL effects should prevail then require of enough 
energy to build up the parton evolution, and the presence of two large and 
similar 
transverse scales. In this work an example of this kind is investigated in 
detail by analytic means: the inclusive hadroproduction of two jets with  
large and similar transverse momenta  and a big relative separation in 
rapidity, the so--called Mueller--Navelet jets. When Y, the distance in 
rapidity between the most forward and backward jets, is not large a fixed 
order perturbative analysis should be enough to describe the cross section but 
when it increases a BFKL resummation of $\left(\alpha_s {\rm Y}\right)^n$ 
terms is needed. 

Mueller--Navelet jets were first proposed in Ref.~\cite{Mueller:1986ey} as 
a clean 
configuration to look for BFKL effects at hadron colliders. A typical 
power--like rise for the partonic cross section was predicted in agreement 
with the value of the asymptotic LL hard Pomeron intercept. However, at 
hadronic level, forward and backward jets are produced in a region of fast 
falling of the parton distributions, reducing the rise of the cross section.  
A way to make small $x$ resummation 
effects more explicit is to look into the azimuthal angle 
decorrelation of the pair of jets. The relevant subprocess  is 
parton + parton $\rightarrow$ jet + jet + any number of soft emissions 
inside the rapidity interval 
separating the two jets. BFKL enhances soft real emission as Y 
increases reducing in this way the amount of angular correlation originally 
present in the back--to--back in transverse plane Born configuration. The LL 
prediction for this azimuthal dependence was first investigated in 
Ref.~\cite{LLResults}.

The results at LL are known to overestimate the rate of decorrelation and 
to lie quite far from the experimental 
data~\cite{Tevatron} as obtained from the Tevatron and subleading higher 
order effects have been 
called for an explanation of this discrepancy. Running coupling effects and 
kinematic constraints have been considered in Ref.~\cite{MNConstraints}. 
In the present work 
the main target is to analytically understand how to include the  
$\alpha_s \left(\alpha_s {\rm Y}\right)^n$ 
next--to--leading logarithmic (NLL) corrections to the BFKL kernel~\cite{FLCC}. 
The effects of this 
kernel were numerically investigated using an 
implementation~\cite{Andersen:2003wy} of the 
NLL iterative solution proposed in Ref.~\cite{Andersen:2003an} 
(different reviews can be found in~\cite{reviews}). It is left for 
future analysis the 
inclusion of the next--to--leading order (NLO) jet vertex~\cite{impactfactors}, the investigation of 
more convergent versions of the kernel~\cite{improvedkernel} and a 
study of parton 
distributions effects in order to make reliable phenomenological predictions at a hadron collider. 
Mueller--Navelet jets should be an important test of our understanding of 
small $x$ resummation to be performed at the Large Hadron Collider at CERN.

After this brief Introduction, in Section 2 the normalization for the gluon 
Green's function is indicated together with the formulae for the 
partonic cross section. Then the operator formalism suggested by Ivanov and Papa 
in Ref.~\cite{Ivanov:2005gn} is 
extended to introduce angular dependences. The form of the NLL kernel for 
all conformal spins calculated by Kotikov and Lipatov in 
Ref.~\cite{Kotikov:2000pm} 
is also discussed. 
Towards the end of the section a compact expression for the angular 
differential cross section 
 which includes the NLL contributions is derived. In Section 3 the numerical 
study of the previous formul{\ae} is discussed in detail. 
Finally, several Conclusions are drawn and different lines for future research 
highlighted.

\section{Calculation of the dijet partonic cross section}

As indicated in the Introduction, in this analysis the object of interest is 
the partonic cross 
section parton + parton $\rightarrow$ jet + jet + soft emission, with the two jets having transverse momenta $\vec{q}_1$ and $\vec{q}_2$ and being produced 
at a large relative rapidity separation Y. This can be related to the external 
hadrons by its approximate relation to the longitudinal momentum fractions
 carried by the jets, {\it i.e.}, ${\rm Y} \sim \ln{x_1 x_2 s/\sqrt{q_1^2 q_2^2}}$, and a convolution with parton distribution functions whose analysis is left for a future work. In the present framework the resummed differential partonic cross section for the particular case of gluon--gluon scattering is
\begin{eqnarray}
\frac{d {\hat \sigma}}{d^2\vec{q}_1 d^2\vec{q}_2} &=& \frac{\pi^2 {\bar \alpha}_s^2}{2} 
\frac{f \left(\vec{q}_1,\vec{q}_2,{\rm Y}\right)}{q_1^2 q_2^2},
\end{eqnarray}
with the usual notation ${\bar \alpha}_s = \alpha_s N_c/\pi$. One can now introduce the Mellin 
transform of the BFKL gluon Green's function in rapidity space:
\begin{eqnarray}
f \left(\vec{q}_1,\vec{q}_2,{\rm Y}\right) &=& \int \frac{d\omega}{2 \pi i} e^{\omega {\rm Y}} f_\omega \left(\vec{q}_1,\vec{q}_2\right).
\end{eqnarray}
The normalization for the BFKL integral equation, including for simplicity only the LL terms, 
then reads 
\begin{eqnarray}
\omega f_\omega \left(\vec{q}_1,\vec{q}_2\right) &=&
\delta^{(2)} \left(\vec{q}_1-\vec{q}_2\right) \nonumber\\
&+&{\bar \alpha}_s \int\frac{d^2 \vec{k}}{\pi \left(\vec{q}_1-\vec{k}\right)^2}
\left(f_\omega \left(\vec{k},\vec{q}_2\right)-\frac{q_1^2 f_\omega \left(\vec{q}_1,\vec{q}_2\right)}{\vec{k}^2+\left(\vec{q}_1-\vec{k}\right)^2}\right).
\label{LLbfkl}
\end{eqnarray}
As it is well known, including the angular dependence on the transverse plane 
of $\vec{q}_1$ and $\vec{q}_2$, the LL solution to Eq.~(\ref{LLbfkl}) can 
be written as
\begin{eqnarray}
f_\omega \left(\vec{q}_1,\vec{q}_2\right) &=& \frac{1}{2 \pi^2} \sum_{n = -\infty}^\infty
\int_{-\infty}^\infty d \nu \,{\left(q_1^2\right)}^{-i \nu -\frac{1}{2}} {\left(q_2^2\right)}^{i \nu -\frac{1}{2}} \frac{e^{i n \left(\theta_1-\theta_2\right)}}{\omega - {\bar \alpha}_s \chi_0 \left(\left|n\right|,\nu\right)} 
\end{eqnarray}
where the eigenvalue of the LL kernel,  
\begin{eqnarray}
\chi_0 \left(n,\nu\right) &=& 2 \psi \left(1\right) - \psi \left(\frac{1}{2}+ i \nu + \frac{n}{2}\right) - \psi\left(\frac{1}{2}- i \nu +\frac{n}{2}\right), 
\end{eqnarray}
is expressed in terms of the logarithmic derivative of the Euler Gamma 
function. As it stands the $n$ variable corresponds to a Fourier transform 
in the angular sector. A more sophisticated interpretation arises if 
Eq.~(\ref{LLbfkl}) is considered for non--zero momentum transfer 
(see, {\it e.g.},~\cite{nf-bfkl}). In this case, if a representation in the 
complex plane for the transverse momenta of the form $\vec{q} = q_x + i q_y$  
is introduced, it can be shown how the BFKL equation corresponds to a 
Schr{\"o}dinger--like equation with a holomorphically separable Hamiltonian 
where $- i \,{\rm Y}$ is the time variable. Both the holomorphic and 
antiholomorphic sectors in the 
Hamiltonian are invariant under spin zero M{\"o}bius transformations with 
eigenfunctions carrying a conformal weight of the form 
$\gamma = \frac{1}{2} + i \nu + \frac{n}{2} $. In the principal series of the 
unitary representation $\nu$ is real and $\left|n\right|$ the integer 
conformal spin~\cite{Lipatov:1985uk}.
 
The partonic cross section is obtained by integration over the 
phase space of the two emitted gluons together with some general jet vertices, {\it i.e.}
\begin{eqnarray}
{\hat \sigma} \left(\alpha_s, {\rm Y},p^2_{1,2}\right) &=&
\int d^2{\vec{q}_1} \int d^2{\vec{q}_2} \,
\Phi_{\rm jet_1}\left(\vec{q}_1,p_1^2\right)
\,\Phi_{\rm jet_2}\left(\vec{q}_2,p_2^2\right)\frac{d {\hat \sigma}}{d^2\vec{q}_1 d^2\vec{q}_2}.
\end{eqnarray}
In the perturbative expansion of these jet vertices,  
$\Phi_{\rm jet_i} = \Phi_{\rm jet_i}^{(0)}+ {\bar \alpha}_s \Phi_{\rm jet_i}^{(1)} + \dots$, only leading--order terms are kept:
\begin{eqnarray}
\Phi_{\rm jet_i}^{(0)} \left(\vec{q},p_i^2\right)&=& \theta \left(q^2-p_i^2\right),
\end{eqnarray}
where $p_i^2$ corresponds to a resolution scale for the transverse momentum of 
the gluon jet. In this way a full NLO accuracy is not 
achieved but it is possible to pin down those effects stemming 
from  the gluon Green's function. To extend this analysis it would be needed to calculate the 
Mellin transform of the NLO jet vertices in Ref.~\cite{impactfactors} where 
the definition of a jet is much more involved than 
here. Therefore one can proceed and write
\begin{eqnarray}
{\hat \sigma} \left(\alpha_s, {\rm Y},p_{1,2}^2\right) =
\frac{\pi^2 {\bar \alpha}_s^2}{2} \int d^2{\vec{q}_1} \int d^2{\vec{q}_2} \,
\frac{\Phi_{\rm jet_1}^{(0)}\left(\vec{q}_1,p_1^2\right)}{q_1^2}
\,\frac{\Phi_{\rm jet_2}^{(0)}\left(\vec{q}_2,p_2^2\right)}{q_2^2}
f \left(\vec{q}_1,\vec{q}_2,{\rm Y}\right). 
\end{eqnarray}
At this stage it is very convenient to recall the work of Ref.~\cite{Ivanov:2005gn} and to 
introduce the following transverse momenta operator representation:
\begin{eqnarray}
{\hat q} \left|\vec{q}_i\right> &=& \vec{q}_i \left|\vec{q}_i\right>
\end{eqnarray}
with the normalization
\begin{eqnarray}
\left<\vec{q}_1\right|{\hat 1}\left|\vec{q}_2\right> &=& \delta^{(2)} \left(\vec{q}_1-\vec{q}_2\right).
\end{eqnarray}
In this notation the BFKL equation simply reads
\begin{eqnarray}
\left(\omega - {\hat K}\right) {\hat f}_\omega &=& {\hat 1} 
\end{eqnarray}
where the kernel has the  expansion
\begin{eqnarray}
{\hat K} &=& {\bar \alpha}_s {\hat K}_0 + {\bar \alpha}_s^2 {\hat K}_1 + \dots
\end{eqnarray}
To NLO accuracy this implies that the solution can be written as
\begin{eqnarray}
{\hat f}_\omega = \left(\omega - {\bar \alpha}_s {\hat K}_0\right)^{-1}
+ {\bar \alpha}_s^2 \left(\omega - {\bar \alpha}_s {\hat K}_0\right)^{-1} 
{\hat K}_1 \left(\omega - {\bar \alpha}_s {\hat K}_0\right)^{-1} + 
{\cal O}\left({\bar \alpha}_s^3\right). 
\label{opGGF}
\end{eqnarray}
The next step is to define a basis on which to express the cross section.
To generalize
 the study in Ref.~\cite{Ivanov:2005gn} this basis should carry not only 
the dependence on the modulus of the transverse momenta but also 
the dependence on their angle on the transverse plane:
\begin{eqnarray}
\left< \vec{q}\right|\left.\nu,n\right> &=& \frac{1}{\pi \sqrt{2}} 
\left(q^2\right)^{i \nu -\frac{1}{2}} \, e^{i n \theta}. 
\label{eignfns}
\end{eqnarray}
The projection $\left<n,\nu\right.\left|\vec{q}\right>$ would be the complex conjugate of 
the previous expression. This basis has been chosen such that it is orthonormal:
\begin{eqnarray}
\left<n',\nu'\right.\left|\nu,n\right>&=& \delta\left(\nu-\nu'\right) 
\delta_{n n'}.
\end{eqnarray}
The action of the NLO kernel on this basis, which was calculated in 
Ref.~\cite{Kotikov:2000pm}, contains non--diagonal terms and can be written as 
\begin{eqnarray}
{\hat K} \left|\nu,n\right> &=& \left\{\frac{}{}{\bar \alpha}_s \, \chi_0\left(\left|n\right|,\nu\right) + {\bar \alpha}_s^2 \, \chi_1\left(\left|n\right|,\nu\right) \right.\nonumber\\
&&\left.\hspace{-2cm}+\,{\bar \alpha}_s^2 \,\frac{\beta_0}{8 N_c}\left[2\,\chi_0\left(\left|n\right|,\nu\right) \left(i \frac{\partial}{\partial \nu}+ \log{\mu^2}\right)+\left(i\frac{\partial}{\partial \nu}\chi_0\left(\left|n\right|,\nu\right)\right)\right]\right\} \left|\nu,n\right>,
\label{opKernel}
\end{eqnarray}
where, from now on, ${\bar \alpha}_s$ stands for ${\bar \alpha}_s \left(\mu^2\right)$, the coupling evaluated at the renormalization point $\mu$ in the $\overline{\rm MS}$ scheme. The first line of Eq.~(\ref{opKernel}) corresponds 
to the scale invariant sector of the kernel. The function $\chi_1$ for a general conformal spin 
reads
\begin{eqnarray}
\chi_1\left(n,\gamma \right) &=& {\cal S} \chi_0 \left(n, \gamma\right)
+ \frac{3}{2} \zeta\left(3\right) - \frac{\beta_0}{8 N_c}\chi_0^2\left(n,\gamma\right)\nonumber\\
&+&\frac{1}{4}\left[\psi''\left(\gamma+\frac{n}{2}\right)+\psi''\left(1-\gamma+\frac{n}{2}\right)-2 \,\phi\left(n,\gamma\right)-2 \,\phi\left(n,1-\gamma\right)\right]\nonumber\\
&-&\frac{\pi^2 \cos{\left(\pi \gamma\right)}}{4 \sin^2\left(\pi \gamma\right)\left(1-2\gamma\right)}\left\{\left[3+\left(1+\frac{n_f}{N_c^3}\right)\frac{2+3\gamma\left(1-\gamma\right)}{\left(3-2\gamma\right)\left(1+2\gamma\right)}\right]\delta_{n 0}\right.\nonumber\\
&&\left.\hspace{2cm}-\left(1+\frac{n_f}{N_c^3}\right)\frac{\gamma\left(1-\gamma\right)}{2\left(3-2\gamma\right)\left(1+2\gamma\right)}\delta_{n 2}\right\}.
\end{eqnarray}
The definitions ${\cal S}=\left(4-\pi^2+5 \beta_0/N_c\right)/12$, and 
$\beta_0 = (11 N_c-2 n_f)/3$, have been used. The function $\phi$ can be 
found in Ref.~\cite{Kotikov:2000pm} and reads
\begin{eqnarray}
\phi(n,\gamma) &=& \sum_{k=0}^{\infty} 
\frac{(-1)^{(k+1)}}{k+\gamma+\frac{n}{2}}
\left(\frac{}{}\psi'(k+n+1)-\psi'(k+1)\right.\nonumber\\
&&\left.\hspace{-2cm}+(-1)^{(k+1)} \left(\beta'(k+n+1)+\beta'(k+1)\right)+\frac{\psi(k+1)-\psi(k+n+1)}{k+\gamma+\frac{n}{2}}\right),
\end{eqnarray}
with
\begin{eqnarray}
4 \,\beta'(\gamma) &=& \psi'\left(\frac{1+\gamma}{2}\right)-\psi'\left(\frac{\gamma}{2}\right).
\end{eqnarray}

In this basis terms with derivatives are associated to the running of the 
coupling~\cite{Forshaw:2000hv}, this is the case in the second line of 
Eq.~(\ref{opKernel}). In particular, the part containing 
$i \frac{\partial \chi_0}{\partial \nu}$, which 
breaks the $\nu \rightarrow -\nu$ symmetry, will be shown to give a 
zero contribution to the cross section when $p_1^2 = p_2^2$. It will also be 
shown how the term with $i \chi_0 \frac{\partial }{\partial \nu}$ mixes in a 
non--trivial way the Green's function with the jet vertices.

To represent the cross section in the present formalism the starting point is 
to project the jet vertices on the basis in Eq.~(\ref{eignfns}):
\begin{eqnarray}
\int d^2{\vec{q}}\,\frac{\Phi_{\rm jet_1}^{(0)}\left(\vec{q},p_1^2\right)}{q^2} \left<\vec{q}\right.\left|\nu,n\right> = 
\frac{1}{\sqrt{2}}\frac{1}{\left(\frac{1}{2}-i \nu\right)}\left(p_1^2\right)^{i \nu - \frac{1}{2}} \delta_{n,0} \equiv c_1\left(\nu\right) \delta_{n,0}.
\label{IFproj}
\end{eqnarray}
The $c_2\left(\nu\right)$ projection of $\Phi_{\rm jet_2}^{(0)}$ on $\left<n,\nu\right|\left.\vec{q}\right>$ is the complex conjugate of~(\ref{IFproj}) with $p_1^2$ being replaced by $p_2^2$. The corresponding inverse relations are
\begin{eqnarray}
\frac{\Phi_{\rm jet_1}^{(0)}\left(\vec{q},p_1^2\right)}{q^2} &=& \sum_{n=-\infty}^{\infty} \int_{-\infty}^\infty d \nu \, c_1\left(\nu\right) \delta_{n,0} \left<n,\nu\right.\left|\vec{q}\right>,\\
\frac{\Phi_{\rm jet_2}^{(0)}\left(\vec{q},p_2^2\right)}{q^2} &=& \sum_{n=-\infty}^{\infty} \int_{-\infty}^\infty  d \nu \, c_2\left(\nu\right) \delta_{n,0} \left<\vec{q}\right.\left|\nu,n\right>.
\end{eqnarray}
The cross section can then be rewritten as
\begin{eqnarray}
{\hat \sigma} \left(\alpha_s, {\rm Y},p_{1,2}^2\right) &=& \frac{\pi^2 {\bar \alpha}_s^2}{2} \sum_{n,n'=-\infty}^\infty \int_{-\infty}^\infty d \nu \int_{-\infty}^\infty d \nu' c_1\left(\nu\right) c_2\left(\nu'\right) \delta_{n,0}
\, \delta_{n',0} \nonumber \\
&&\hspace{1cm} \times \int \frac{d \omega}{2 \pi i} \, e^{\omega {\rm Y}} \left<n,\nu\right|{\hat f}_\omega \left|\nu',n'\right>.
\end{eqnarray}
Making use of the operator representation in Eq.~(\ref{opGGF}), the action 
of the kernel in Eq.~(\ref{opKernel}) and integration by parts, ${\hat \sigma}$ can be expressed as
\begin{eqnarray}
{\hat \sigma} \left(\alpha_s, {\rm Y},p_{1,2}^2\right) &=&
\frac{\pi^2 {\bar \alpha}_s^2}{2} \sum_{n=-\infty}^\infty \int_{-\infty}^\infty d \nu 
\,e^{{\bar \alpha}_s \chi_0\left(\left|n\right|,\nu\right) {\rm Y}} 
c_1\left(\nu\right) c_2\left(\nu\right) \delta_{n,0} \label{logdercross}\\
&&\hspace{-3.3cm}\times \left\{1+{\bar \alpha}_s^2 \, {\rm Y} \left[\chi_1\left(\left|n\right|,\nu\right)+\frac{\beta_0}{4 N_c} \left(\log{(\mu^2)}+ \frac{i}{2} \frac{\partial}{\partial \nu}\log{\left(\frac{c_1\left(\nu\right)}{c_2\left(\nu\right)}\right)}+ \frac{i}{2} \frac{\partial}{\partial \nu}\right)\chi_0\left(\left|n\right|,\nu\right)\right]\right\}.\nonumber
\end{eqnarray}
For the LO jet vertices the logarithmic derivative in 
Eq.~(\ref{logdercross}) explicitly reads
\begin{eqnarray}
- i \frac{\partial}{\partial \nu}\log{\left(\frac{c_1\left(\nu\right)}{c_2\left(\nu\right)}\right)} &=& \log{\left(p_1^2p_2^2\right)}+ \frac{1}{\frac{1}{4}+\nu^2}.
\end{eqnarray}
The angular differential cross section can be calculated considering the 
following representation of the $c_1 c_2$ product:
\begin{eqnarray}
c_1(\nu) \,c_2(\nu) \,\delta_{n,0}&=& \frac{1}{2 \sqrt{p_1^2 p_2^2}} \frac{1}{\left(\frac{1}{4}+\nu^2\right)}\left(\frac{p_1^2}{p_2^2}\right)^{i \nu} 
\int_{-\pi}^\pi \frac{d \phi}{2 \pi}  e^{i n \phi}, 
\end{eqnarray}
with $\phi = \theta_1-\theta_2 - \pi$. 
Therefore, in the case where the two resolution momenta are equal, 
$p_1^2 = p_2^2 \equiv p^2$, the angular differential cross section can 
be expressed as
\begin{eqnarray}
\frac{d {\hat \sigma}\left(\alpha_s, {\rm Y},p^2\right)}{d \phi}  &=&
\frac{\pi^2 {\bar \alpha}_s^2}{4 p^2} \sum_{n=-\infty}^\infty 
\frac{1}{2 \pi}e^{i n \phi} \int_{-\infty}^\infty d \nu 
\,e^{{\bar \alpha}_s \chi_0\left(\left|n\right|,\nu\right) {\rm Y}} 
\frac{1}{\left(\frac{1}{4}+\nu^2\right)}\nonumber\\
&&\hspace{-3cm}\times \left\{1+{\bar \alpha}_s^2 \, {\rm Y} \left[\chi_1\left(\left|n\right|,\nu\right)-\frac{\beta_0}{8 N_c} \chi_0\left(\left|n\right|,\nu\right) \left(2 \log{\left(\frac{p^2}{\mu^2}\right)}+\frac{1}{\left(\frac{1}{4}+\nu^2\right)}\right)\right]\right\}.
\label{Noexp}
\end{eqnarray}
The term proportional to $\frac{\partial \chi_0}{\partial \nu}$ in 
Eq.~(\ref{logdercross}) gives no contribution after 
integration as it is an odd function in $\nu$. Within NLO accuracy there is 
freedom to exponentiate the integrand of this result. In this work this is 
done both for the scale invariant and for the scale dependent terms, in close 
resemblance with the property of reggeization. Furthermore, renormalization 
group improved perturbation theory is called for to introduce the replacement
\begin{eqnarray}
{\bar \alpha}_s - {\bar \alpha}_s^2 \frac{\beta_0}{4 N_c} \log{\left(\frac{p^2}{\mu^2}\right)} &\rightarrow& {\bar \alpha}_s 
\left(p^2\right).
\end{eqnarray}
In this way the differential 
distribution can be conveniently rewritten as
\begin{eqnarray}
\frac{d {\hat \sigma}\left(\alpha_s, {\rm Y},p^2\right)}{d \phi}  &=&
\frac{\pi^3 {\bar \alpha}_s^2}{2 p^2} \frac{1}{2 \pi}\sum_{n=-\infty}^\infty 
e^{i n \phi} {\cal C}_n \left({\rm Y}\right),
\end{eqnarray}
with
\begin{eqnarray}
{\cal C}_n \left({\rm Y}\right) =
\int_{-\infty}^\infty \frac{d \nu}{2 \pi}\frac{e^{{\bar \alpha}_s \left(p^2\right){\rm Y} \left(\chi_0\left(\left|n\right|,\nu\right)+{\bar \alpha}_s  \left(p^2\right) \left(\chi_1\left(\left|n\right|,\nu\right)-\frac{\beta_0}{8 N_c} \frac{\chi_0\left(\left|n\right|,\nu\right)}{\left(\frac{1}{4}+\nu^2\right)}\right)\right)}}{\left(\frac{1}{4}+\nu^2\right)}.
\label{Cn}
\end{eqnarray}
Different interesting observables can be constructed with these coefficients 
and they will be studied in detail in the next section. Due to the large and 
negative, with respect to the LL terms, size of the NLL corrections it will 
turn out that the exponentiated form in Eq.~(\ref{Cn}) is mandatory in order 
to reach convergent results. This is discussed below.

\section{Analysis of convergence and study of observables}

To investigate the properties of the different parts of the kernel it is 
useful to introduce five types of coefficients. The first one is
\begin{eqnarray}
{\cal C}_n^{\rm LL} \left({\rm Y}\right) &=&
\int_{-\infty}^\infty \frac{d \nu}{2\pi} 
\frac{e^{{\bar \alpha}_s {\rm Y}\chi_0\left(\left|n\right|,\nu\right)}}{\left(\frac{1}{4}+\nu^2\right)},
\end{eqnarray}
defined at LL accuracy and previously studied in the literature (see last 
reference in~\cite{LLResults}). When all the NLL terms are exponentiated as in 
Eq.~(\ref{Cn}) it will be referred to as ${\cal C}_n^{\rm NLL}$, while if the 
NLL pieces are not exponentiated, as in Eq.~(\ref{Noexp}), it is named  
${\cal C}_n^{\rm Expansion}$. If in Eq.~(\ref{Cn}) the $\chi_1$ kernel is 
removed then the coefficients correspond to the case of LL plus running 
coupling and it is noted as ${\cal C}_n^{\rm Running\,coupling}$. Finally, 
the scale invariant contributions can be isolated by setting $\beta_0$ to zero 
in the exponent of Eq.~(\ref{Cn}) and this will be called 
${\cal C}_n^{\rm Scale\,invariant}$.

\begin{figure}[tbp]
\epsfig{width=8cm,file=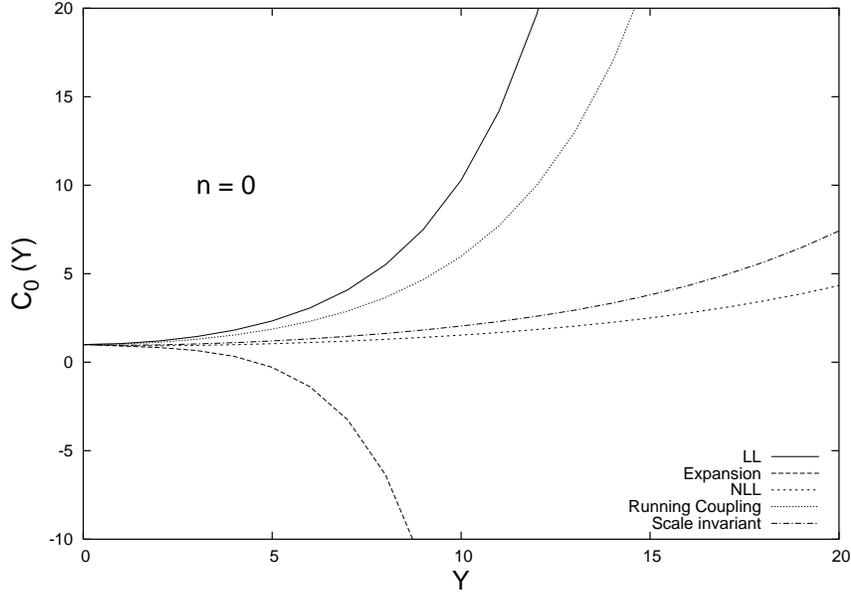,angle=-90}
\caption{Evolution in Y of the ${\cal C}_0 ({\rm Y})$ coefficient.}
\label{CnvsY0}
\end{figure}

The coefficient governing the energy dependence of the cross section 
corresponds to $n=0$:
\begin{eqnarray}
{\hat \sigma}\left(\alpha_s, {\rm Y},p^2\right) &=& 
\frac{\pi^3 {\bar \alpha}_s^2}{2 p^2} \, {\cal C}_0 \left({\rm Y}\right).
\end{eqnarray}
As the NLL corrections are large and negative, when not exponentiated 
they lead to a non--convergent behavior. This is seen in Fig.~\ref{CnvsY0} 
where the resolution scale $p = 30 \, {\rm GeV}$ has been chosen. The 
values $n_f = 4$ and $\Lambda_{\rm QCD} = 0.1416$ GeV were taken in 
${\bar \alpha}_s \left(p^2\right) = 4 N_c / \left(\beta_0 
\ln{\left(p^2/\Lambda_{\rm QCD}^2\right)}\right)$. In this plot it can be 
seen how the coefficient ${\cal C}_0^{\rm Expansion}$ generates an unphysical 
behaviour in contrast to the exponential rise associated to the other 
coefficients. Nevertheless it should be noticed that the term proportional to 
${\bar \alpha_s}^2 {\rm Y}$
\begin{eqnarray}
\chi_0  \left(\frac{c_1^{(1)}}{c_1^{(0)}}+\frac{c_2^{(1)}}{c_2^{(0)}}\right)
\end{eqnarray}
would be added to $\chi_1$ in Eq.~(\ref{Noexp}) if the NLO jet vertex 
$c^{(0)} + {\bar \alpha_s} c^{(1)}$ was  
brought into the calculation. The convergence properties of Eq.~(\ref{Noexp}) 
might well change in that case. 

A familiar consequence of introducing the effects of the running of the 
coupling is that the LL intercept is reduced as can be seen also in 
Fig.~\ref{CnvsY0}. Meanwhile, if the scale 
invariant sector of the NLL kernel, $\chi_1$, is also introduced then a further decrease of this rise takes place. The $n = 0$ coefficient is directly related 
to the normalized cross section
\begin{eqnarray}
\frac{{\hat \sigma} \left({\rm Y}\right)}{{\hat \sigma} \left(0\right)} 
&=& \frac{{\cal C}_0 \left({\rm Y}\right)}{{\cal C}_0\left(0\right)}. 
\end{eqnarray}

\begin{figure}[tbp]
\centering
  \epsfig{width=8cm,file=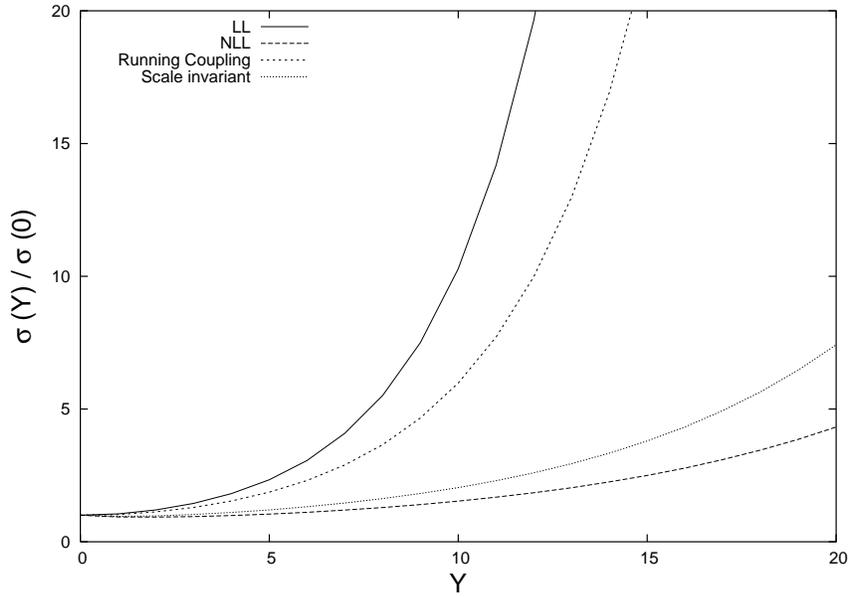,angle=-90}
\caption{Evolution of the partonic cross section with the rapidity separation of the dijets. }
\label{SectionconY}
\end{figure}

\begin{figure}[tbp]
\centering
\hspace{0.4cm}
\epsfig{width=8cm,file=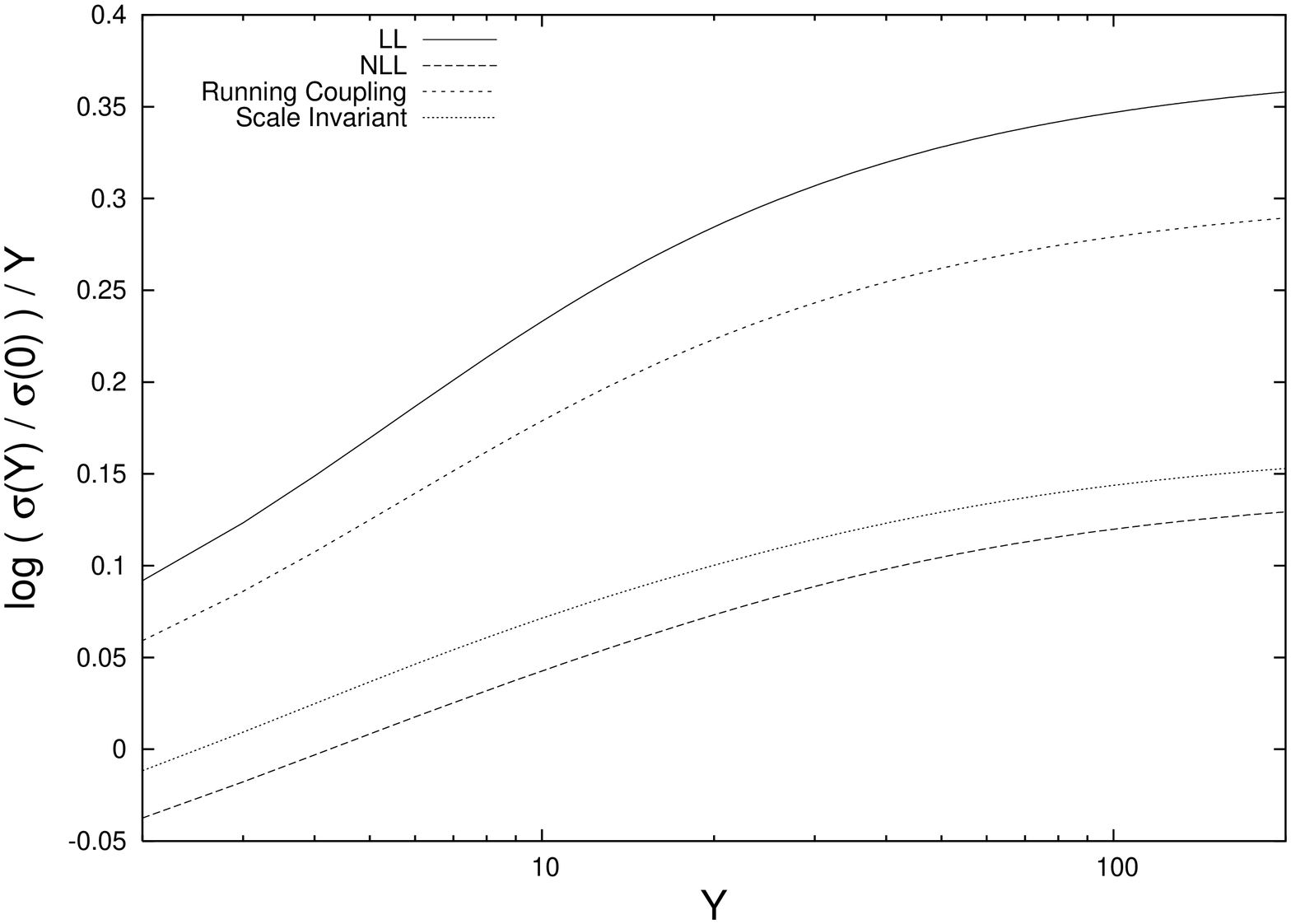,angle=-90}
\caption{Intercepts for the cross section as a function of rapidity.}
\label{Inter}
\end{figure}

The rise in rapidity of this observable is shown in 
Fig.~\ref{SectionconY}. Clearly the NLL intercept is very much reduced with 
respect to the LL case. In Fig.~\ref{Inter} large values of Y are 
considered in order to approach the asymptotic regime. The LL intercept 
tends to the asymptotic value of $4 \,{\bar \alpha}_s (30) \ln{2} \sim 0.37$ 
while the NLL result lies around one third of this number.

\begin{figure}[tbp]
\centering
\hspace{0.4cm}
\epsfig{width=6cm,file=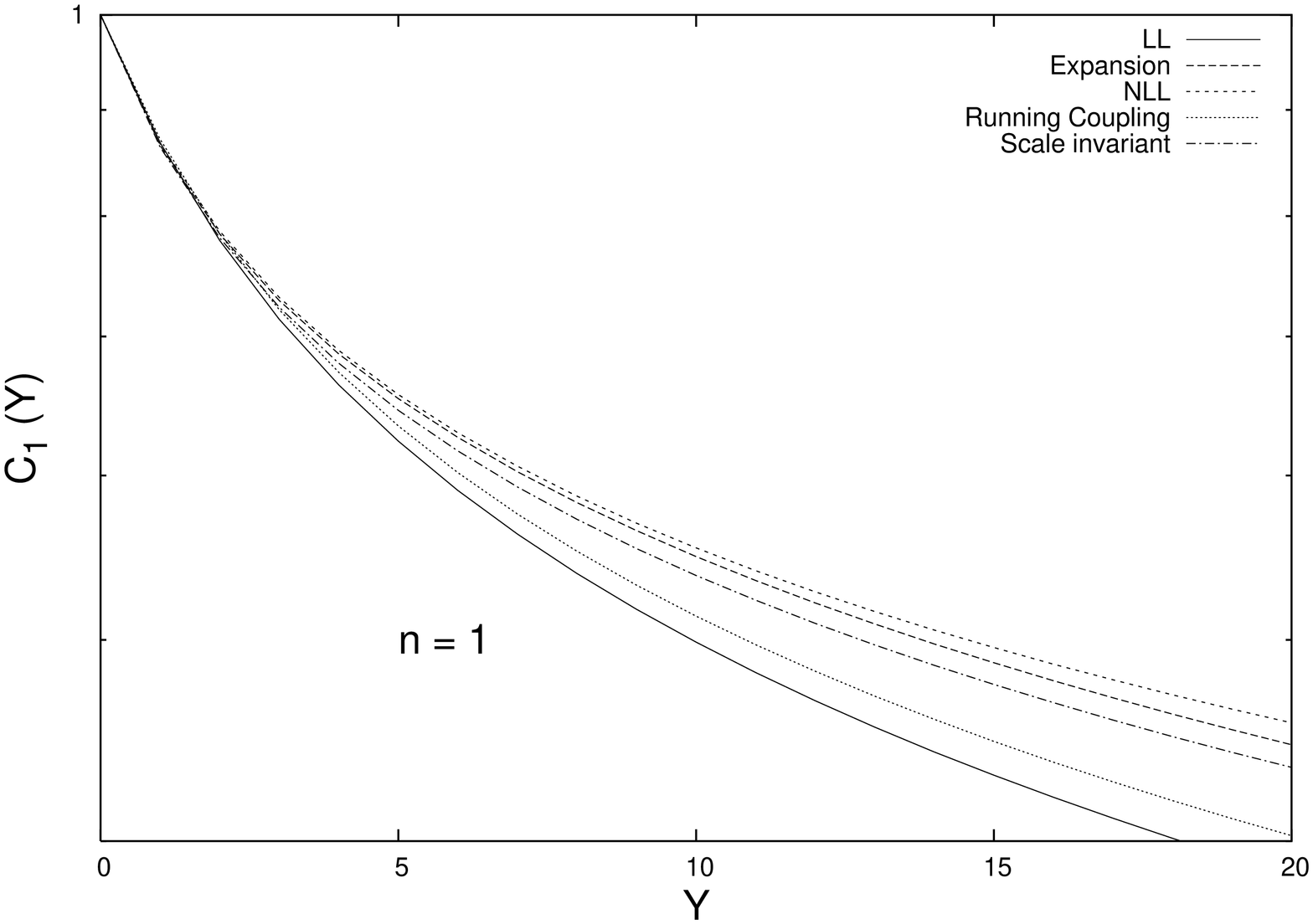,angle=-90}
\epsfig{width=6cm,file=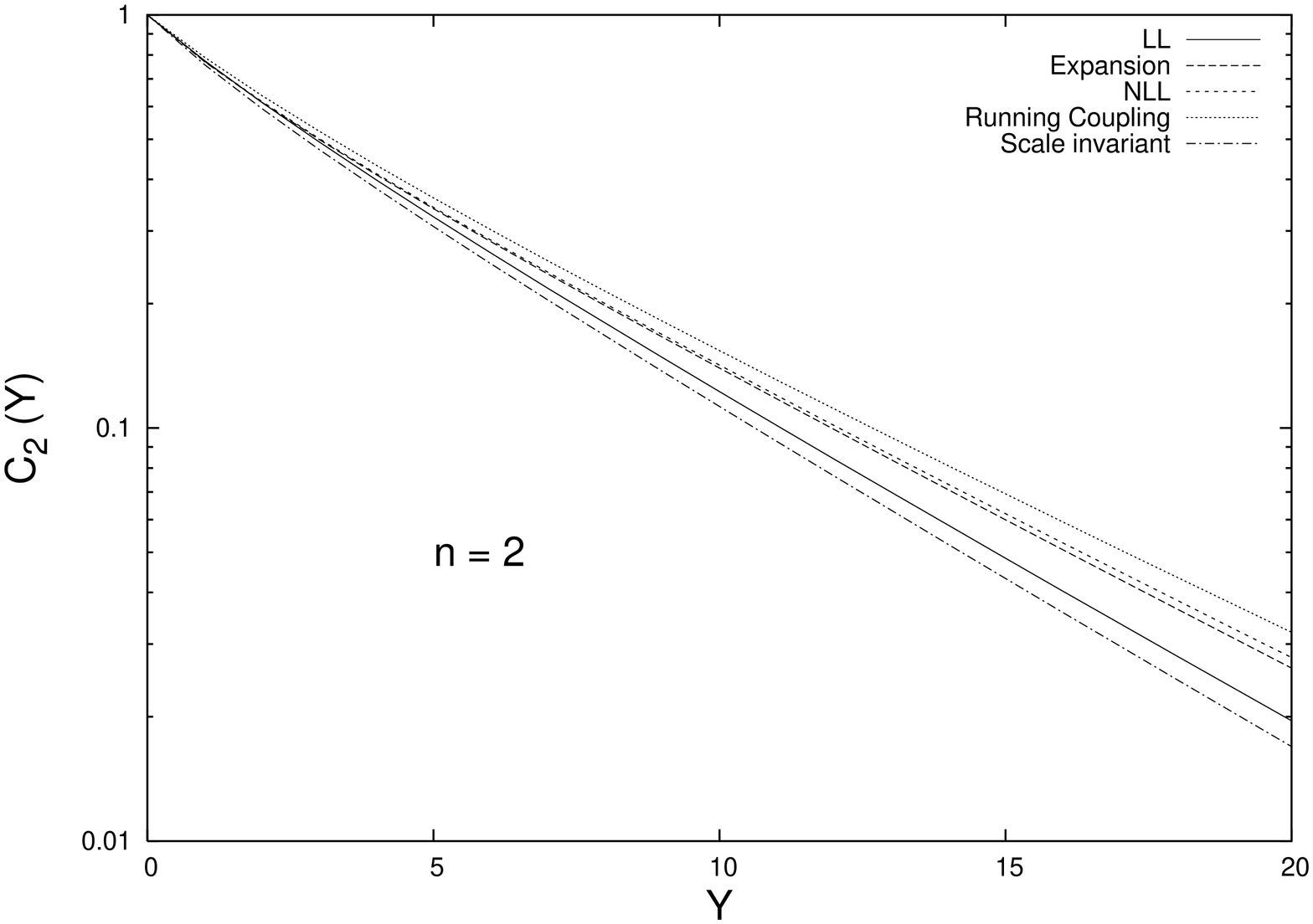,angle=-90}
\epsfig{width=6cm,file=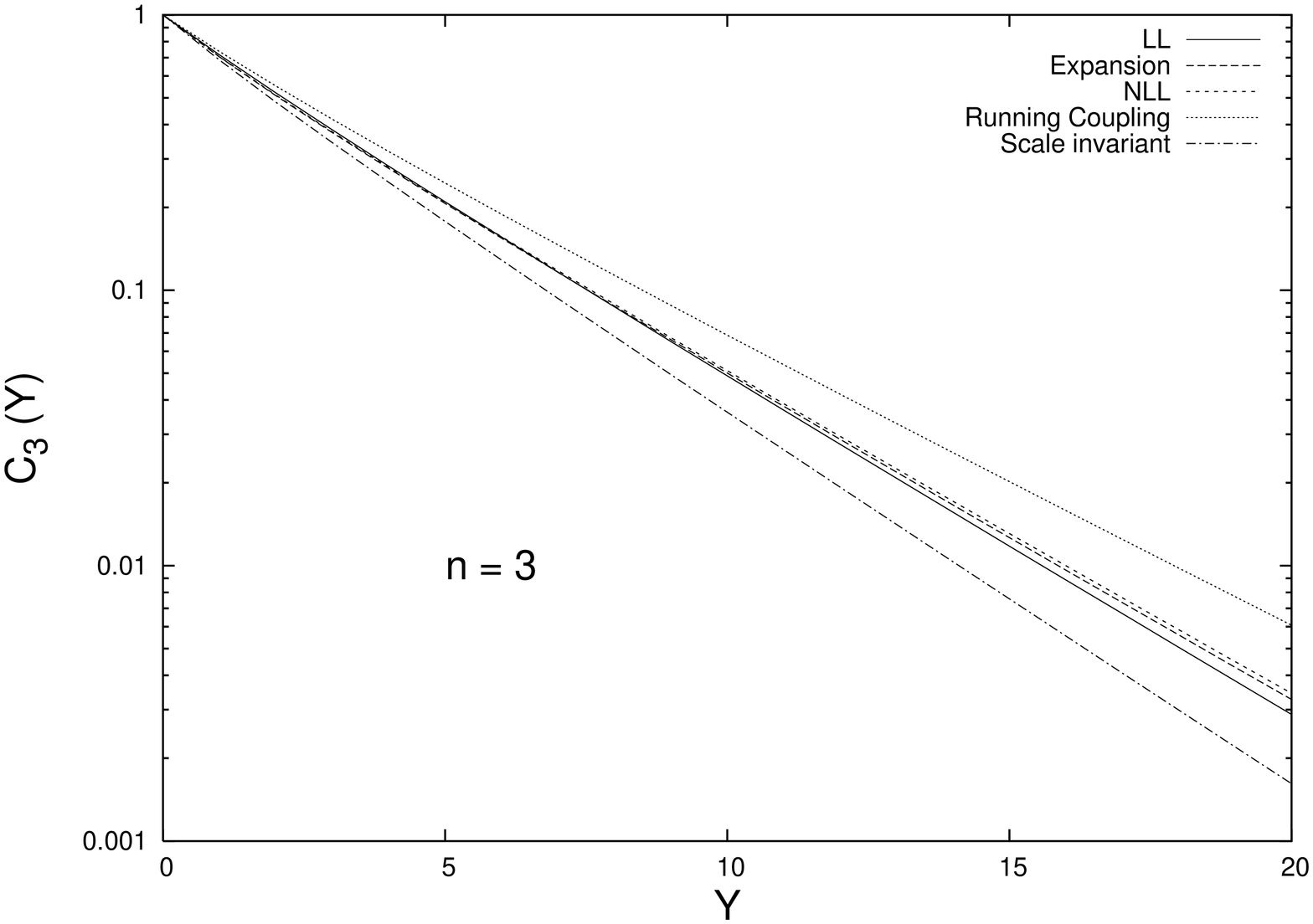,angle=-90}
\caption{Evolution in Y of the ${\cal C}_n ({\rm Y})$ coefficients for $n=1,2,3$.}
\label{CnvsY123}
\end{figure}

\begin{figure}[tbp]
\centering
  \epsfig{width=8cm,file=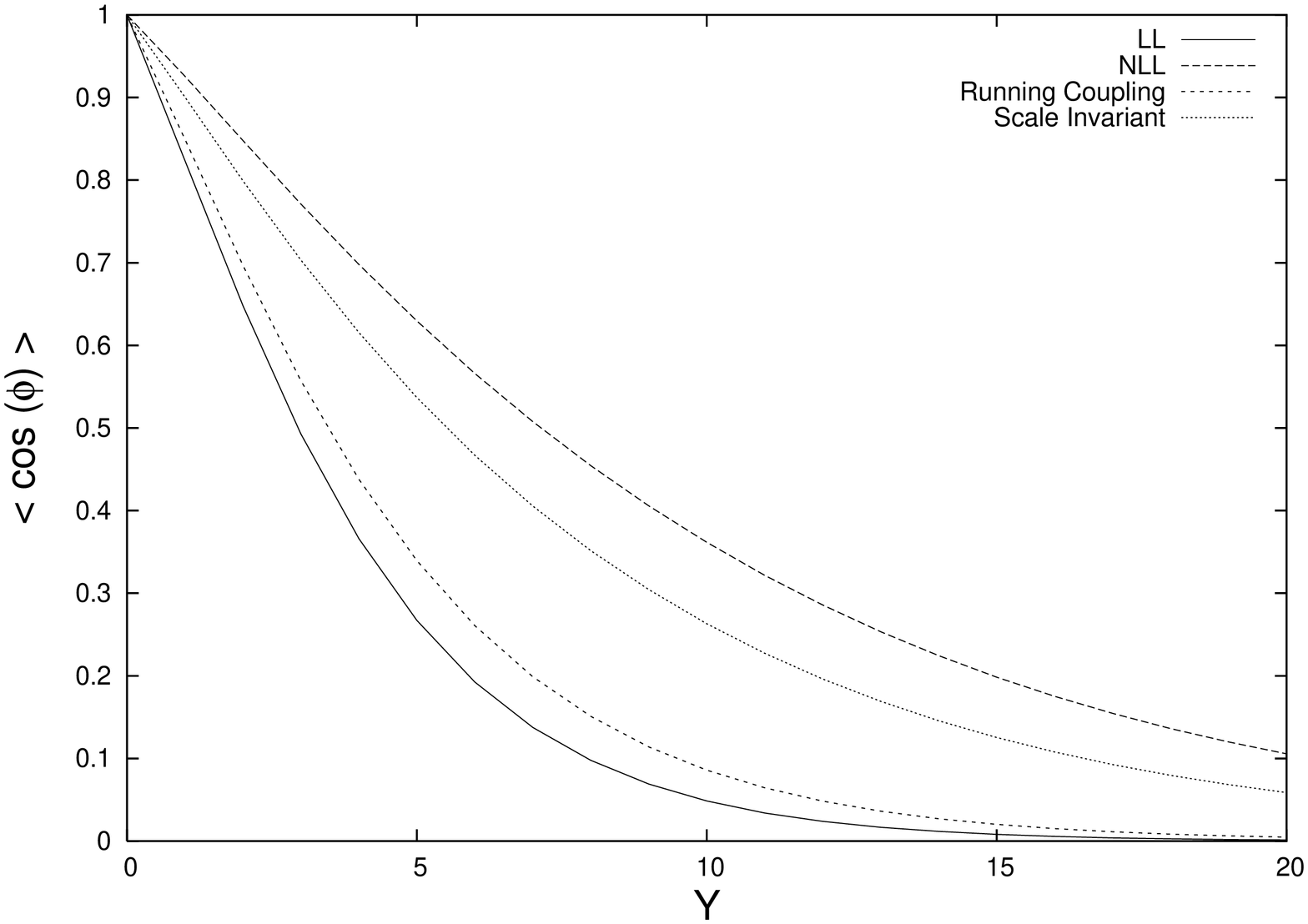 ,angle=-90}
\caption{Dijet azimuthal angle decorrelation as a function of their separation in rapidity.}
\label{Cos1Y}
\end{figure}

\begin{figure}[tbp]
\centering
  \epsfig{width=8cm,file=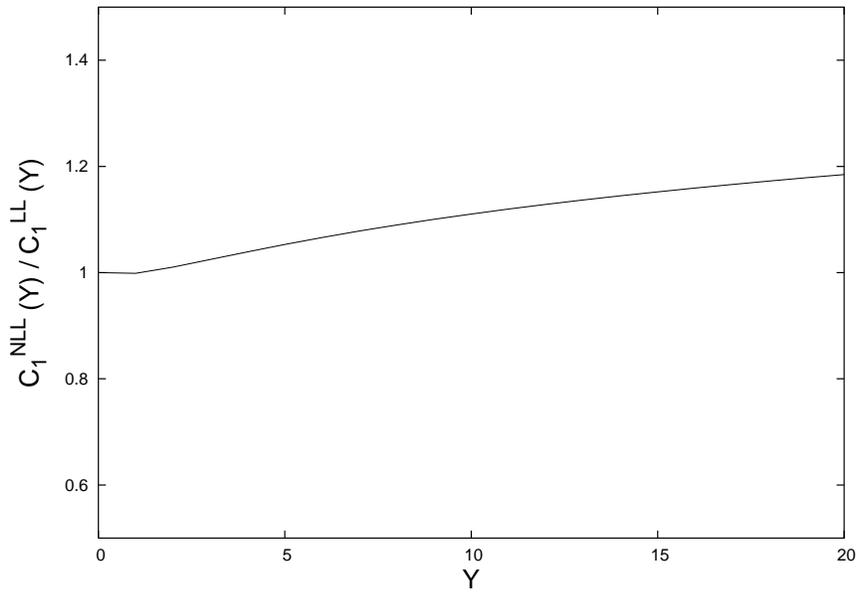 ,angle=-90}
\caption{Comparative ratio between the NLL and LL coefficients for $n=1$ conformal 
spin.}
\label{CosRatio}
\end{figure}

\begin{figure}[tbp]
\centering
  \epsfig{width=8cm,file=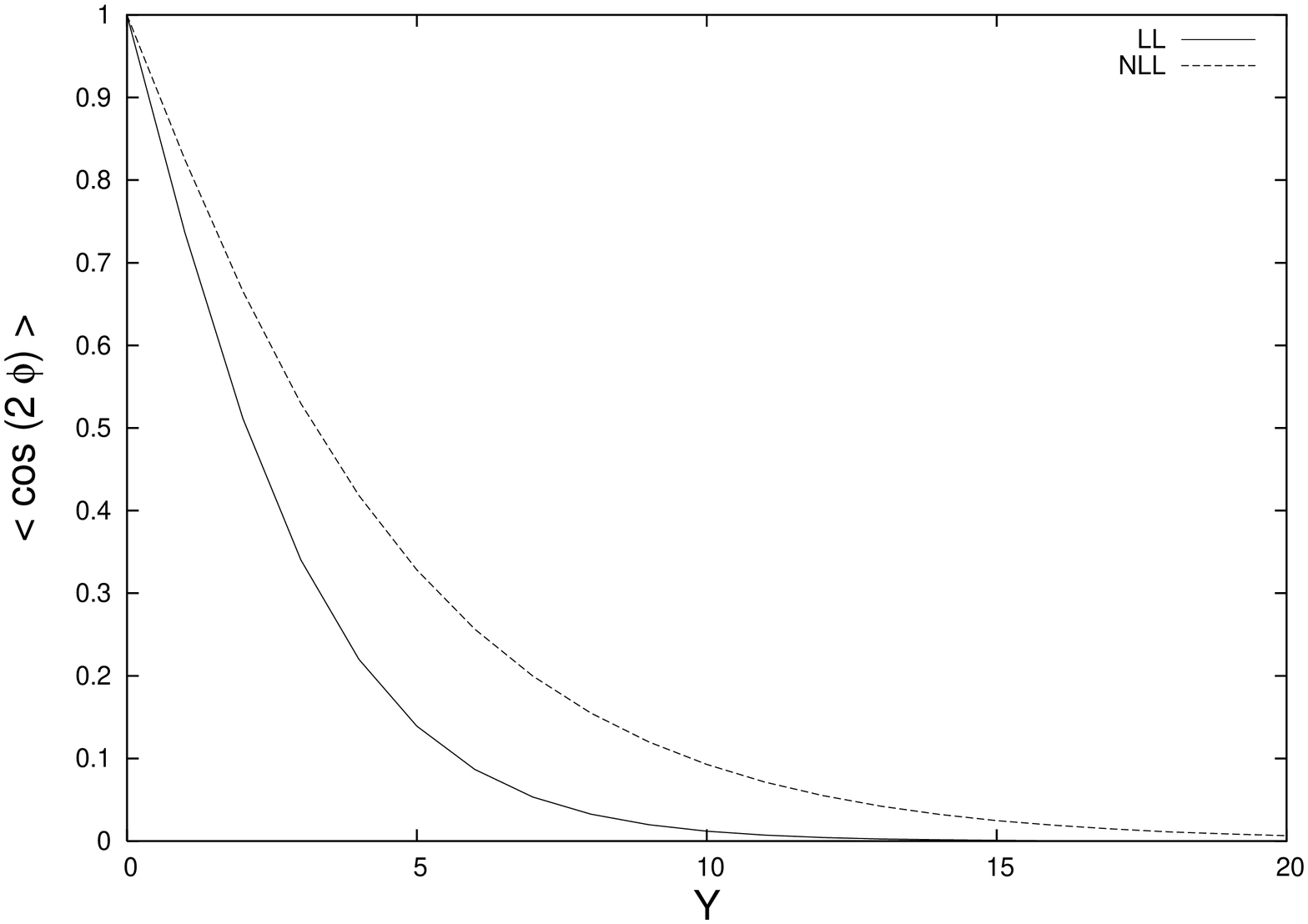 ,angle=-90}
  \epsfig{width=8cm,file=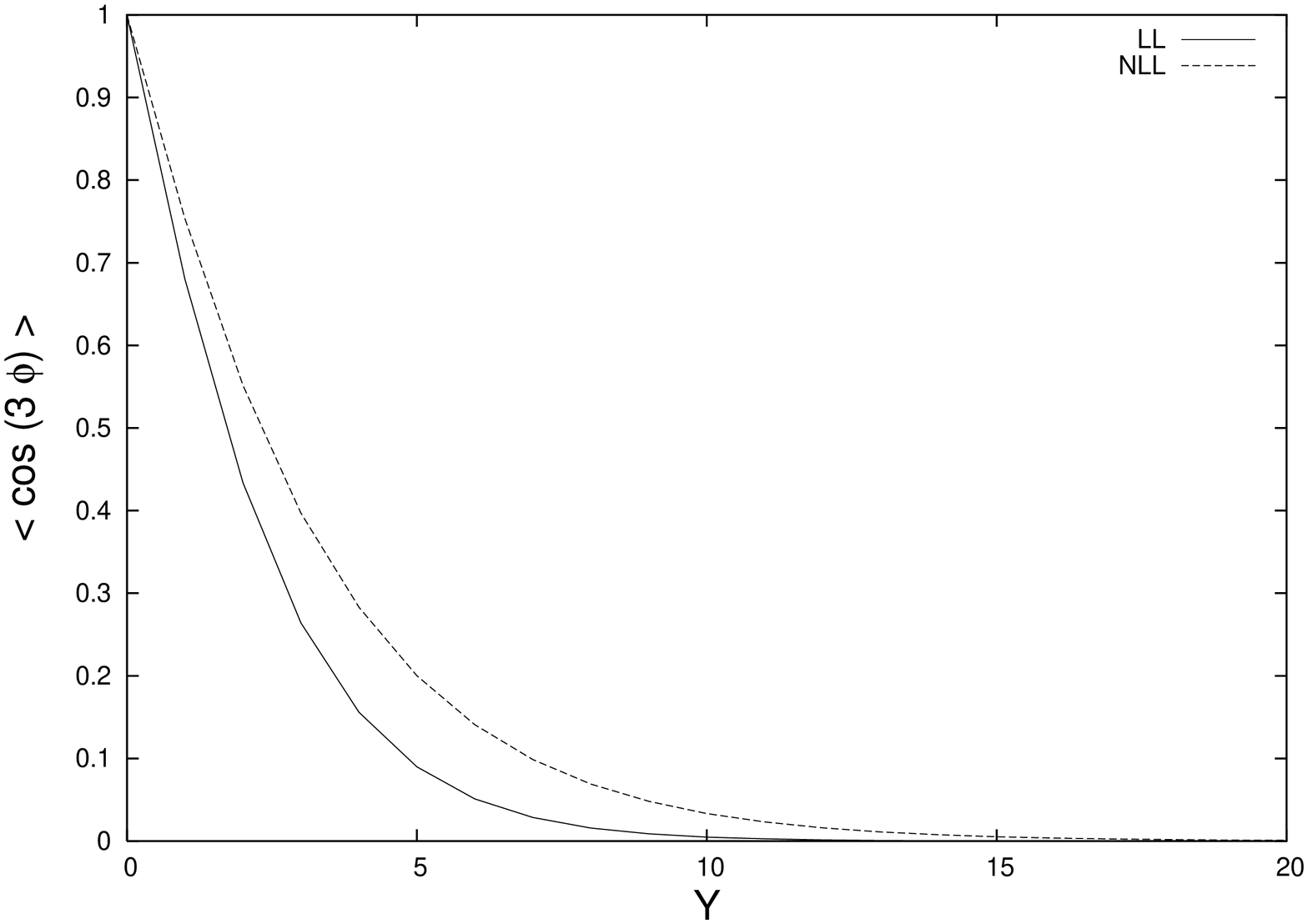 ,angle=-90}
\caption{Dijet azimuthal angle decorrelation as a function of their separation in rapidity.}
\label{Cos23Y}
\end{figure}

Independently of the NLL kernel being exponentiated or not, the remaining 
coefficients with $n \geq 1$ all decrease with Y. This can be seen in the 
plots of Fig.~\ref{CnvsY123}.
The consequence of this decrease is that the angular correlations also 
diminish as the rapidity interval between the jets gets larger. This point 
can be studied in detail using the mean values 
\begin{eqnarray}
\left<\cos{\left( m \phi \right)}\right> &=& \frac{{\cal C}_m \left({\rm Y}\right)}{{\cal C}_0\left({\rm Y}\right)}.
\end{eqnarray}
$\left<\cos{\left(\phi\right)}\right>$ is calculated in Fig.~\ref{Cos1Y}. The 
most important consequence of this plot is that the NLL effects dramatically 
decrease the azimuthal angle decorrelation. This is already the case when 
only the running of the coupling is introduced but the scale invariant terms 
make this effect much bigger. This is encouraging from the 
phenomenological point of view given that the data at the Tevatron typically 
have lower decorrelation than predicted by LL BFKL or LL with running coupling. 
It is worth noting that the 
difference in the prediction for decorrelation between 
LL and NLL is mostly driven by the $n=0$ conformal spin. This can be 
understood looking at the ratio
\begin{eqnarray}
\frac{\left<\cos{\left(\phi \right)}\right>^{\rm NLL}}{\left<\cos{\left(\phi \right)}\right>^{\rm LL}} &=& \frac{{\cal C}_1^{\rm NLL} \left({\rm Y}\right)}{{\cal C}_0^{\rm NLL}\left({\rm Y}\right)}\frac{{\cal C}_0^{\rm LL} \left({\rm Y}\right)}{{\cal C}_1^{\rm LL}\left({\rm Y}\right)},
\end{eqnarray}
and noticing that 
\begin{eqnarray}
1.2 > \frac{{\cal C}_1^{\rm NLL} \left({\rm Y}\right)}{{\cal C}_1^{\rm LL}\left({\rm Y}\right)} > 1.
\end{eqnarray}
This ratio is calculated in Fig.~\ref{CosRatio}.
This point is a consequence of the good convergence in terms of asymptotic 
intercepts of the NLL BFKL calculation for conformal spins larger than zero. 
In particular the $n=1$ case is special in that the property of zero intercept 
at LL, $\chi_0(1,1/2) = 0$, is preserved under radiative corrections since
\begin{eqnarray}
\chi_1\left(1,\frac{1}{2}\right) = 
{\cal S} \chi_0 \left(1, \frac{1}{2}\right)
+ \frac{3}{2} \zeta\left(3\right) 
- \frac{\beta_0}{8 N_c}\chi_0^2\left(1, \frac{1}{2}\right)
+\frac{\psi''\left(1\right)}{2} - \phi\left(1, \frac{1}{2}\right)
\end{eqnarray}
is also zero. 
For completeness the $m=2,3$ cases for $\left<\cos{\left(m \phi\right)}\right>$ 
are shown in Fig.~\ref{Cos23Y}. 
These distributions are relevant because they prove the structure of the 
higher conformal spins. The trend is the same as previously discussed: the 
correlation increases as higher order corrections in the small $x$ resummation 
are included. 

\section{Conclusions}

An analytic procedure has been presented to calculate the effect of higher order 
corrections in the description of Mueller--Navelet 
jets where two jets with moderately high and similar transverse momentum
 are produced at a large 
relative rapidity separation in hadron--hadron collisions. This is a 
promising observable to study small $x$ physics at the Large Hadron Collider 
at CERN given its large energy range. The focus of 
the analysis has been on those effects with direct origin in the NLO BFKL kernel, 
while the jet vertices have been considered at LO accuracy. It 
has been shown how the growth with energy of the cross section is reduced 
when going from a LL to a NLL approximation, 
and how the azimuthal angle decorrelations largely decrease due to the 
higher order effects. The present study has been performed at partonic 
level while the implementation of a full analysis, including parton 
distribution 
functions, NLO jet vertices and the investigation of 
collinearly improved kernels, will be published elsewhere. 

\begin{flushleft}
{\bf \large Acknowledgments}
\end{flushleft}
The author acknowledges interesting discussions with G Altarelli, J R Andersen, 
V Del Duca, J R Forshaw, D Ivanov, A Kotikov, G P Salam, M H Seymour and, 
in particular, J Bartels, L Lipatov, A Papa and F Schwennsen.

\end{document}